\documentclass[aps,prd,twocolumn,superscriptaddress,showkeys,amsmath,amssymb]{revtex4} 

\usepackage{graphicx}
\usepackage{dcolumn}
\usepackage{setspace}
\usepackage{subfigure}
\usepackage{slashbox}
\usepackage{epsf}
\usepackage{epsfig}

\begin{document}



\newcommand{\mtop}{\mbox{$M_{\rm top}$}}
\newcommand{\ttbar}{\mbox{$t\bar{t}$}}
\newcommand{\thetastar}{\mbox{$\theta^*$}}
\newcommand{\costhetastar}{\mbox{$\cos\thetastar$}}
\newcommand{\Mlb}{\mbox{$M_{lb}$}}
\newcommand{\Et}{\mbox{$E_T$}}
\newcommand{\Pt}{\mbox{$p_T$}}
\newcommand{\Flong}{\mbox{$F_0$}}
\newcommand{\Fplus}{\mbox{$F_+$}}
\newcommand{\met}{\mbox{$\protect \raisebox{0.3ex}{$\not$}\Et$}}

\bibliographystyle{revtex}

\vspace*{1.5cm}

\title{\boldmath 
        Measurement of the  Top Quark Mass in the All-Hadronic Mode at CDF}

\affiliation{Institute of Physics, Academia Sinica, Taipei, Taiwan 11529, Republic of China}
\affiliation{Argonne National Laboratory, Argonne, Illinois 60439, USA}
\affiliation{University of Athens, 157 71 Athens, Greece}
\affiliation{Institut de Fisica d'Altes Energies, ICREA, Universitat Autonoma de Barcelona, E-08193, Bellaterra (Barcelona), Spain}
\affiliation{Baylor University, Waco, Texas 76798, USA}
\affiliation{Istituto Nazionale di Fisica Nucleare Bologna, $^{ee}$University of Bologna, I-40127 Bologna, Italy}
\affiliation{University of California, Davis, Davis, California 95616, USA}
\affiliation{University of California, Los Angeles, Los Angeles, California 90024, USA}
\affiliation{Instituto de Fisica de Cantabria, CSIC-University of Cantabria, 39005 Santander, Spain}
\affiliation{Carnegie Mellon University, Pittsburgh, Pennsylvania 15213, USA}
\affiliation{Enrico Fermi Institute, University of Chicago, Chicago, Illinois 60637, USA}
\affiliation{Comenius University, 842 48 Bratislava, Slovakia; Institute of Experimental Physics, 040 01 Kosice, Slovakia}
\affiliation{Joint Institute for Nuclear Research, RU-141980 Dubna, Russia}
\affiliation{Duke University, Durham, North Carolina 27708, USA}
\affiliation{Fermi National Accelerator Laboratory, Batavia, Illinois 60510, USA}
\affiliation{University of Florida, Gainesville, Florida 32611, USA}
\affiliation{Laboratori Nazionali di Frascati, Istituto Nazionale di Fisica Nucleare, I-00044 Frascati, Italy}
\affiliation{University of Geneva, CH-1211 Geneva 4, Switzerland}
\affiliation{Glasgow University, Glasgow G12 8QQ, United Kingdom}
\affiliation{Harvard University, Cambridge, Massachusetts 02138, USA}
\affiliation{Division of High Energy Physics, Department of Physics, University of Helsinki and Helsinki Institute of Physics, FIN-00014, Helsinki, Finland}
\affiliation{University of Illinois, Urbana, Illinois 61801, USA}
\affiliation{The Johns Hopkins University, Baltimore, Maryland 21218, USA}
\affiliation{Institut f\"{u}r Experimentelle Kernphysik, Karlsruhe Institute of Technology, D-76131 Karlsruhe, Germany}
\affiliation{Center for High Energy Physics: Kyungpook National University, Daegu 702-701, Korea; Seoul National University, Seoul 151-742, Korea; Sungkyunkwan University, Suwon 440-746, Korea; Korea Institute of Science and Technology Information, Daejeon 305-806, Korea; Chonnam National University, Gwangju 500-757, Korea; Chonbuk National University, Jeonju 561-756, Korea}
\affiliation{Ernest Orlando Lawrence Berkeley National Laboratory, Berkeley, California 94720, USA}
\affiliation{University of Liverpool, Liverpool L69 7ZE, United Kingdom}
\affiliation{University College London, London WC1E 6BT, United Kingdom}
\affiliation{Centro de Investigaciones Energeticas Medioambientales y Tecnologicas, E-28040 Madrid, Spain}
\affiliation{Massachusetts Institute of Technology, Cambridge, Massachusetts 02139, USA}
\affiliation{Institute of Particle Physics: McGill University, Montr\'{e}al, Qu\'{e}bec, Canada H3A~2T8; Simon Fraser University, Burnaby, British Columbia, Canada V5A~1S6; University of Toronto, Toronto, Ontario, Canada M5S~1A7; and TRIUMF, Vancouver, British Columbia, Canada V6T~2A3}
\affiliation{University of Michigan, Ann Arbor, Michigan 48109, USA}
\affiliation{Michigan State University, East Lansing, Michigan 48824, USA}
\affiliation{Institution for Theoretical and Experimental Physics, ITEP, Moscow 117259, Russia}
\affiliation{University of New Mexico, Albuquerque, New Mexico 87131, USA}
\affiliation{The Ohio State University, Columbus, Ohio 43210, USA}
\affiliation{Okayama University, Okayama 700-8530, Japan}
\affiliation{Osaka City University, Osaka 588, Japan}
\affiliation{University of Oxford, Oxford OX1 3RH, United Kingdom}
\affiliation{Istituto Nazionale di Fisica Nucleare, Sezione di Padova-Trento, $^{ff}$University of Padova, I-35131 Padova, Italy}
\affiliation{University of Pennsylvania, Philadelphia, Pennsylvania 19104, USA}
\affiliation{Istituto Nazionale di Fisica Nucleare Pisa, $^{gg}$University of Pisa, $^{hh}$University of Siena and $^{ii}$Scuola Normale Superiore, I-56127 Pisa, Italy}
\affiliation{University of Pittsburgh, Pittsburgh, Pennsylvania 15260, USA}
\affiliation{Purdue University, West Lafayette, Indiana 47907, USA}
\affiliation{University of Rochester, Rochester, New York 14627, USA}
\affiliation{The Rockefeller University, New York, New York 10065, USA}
\affiliation{Istituto Nazionale di Fisica Nucleare, Sezione di Roma 1, $^{jj}$Sapienza Universit\`{a} di Roma, I-00185 Roma, Italy}
\affiliation{Rutgers University, Piscataway, New Jersey 08855, USA}
\affiliation{Texas A\&M University, College Station, Texas 77843, USA}
\affiliation{Istituto Nazionale di Fisica Nucleare Trieste/Udine, I-34100 Trieste, $^{kk}$University of Udine, I-33100 Udine, Italy}
\affiliation{University of Tsukuba, Tsukuba, Ibaraki 305, Japan}
\affiliation{Tufts University, Medford, Massachusetts 02155, USA}
\affiliation{University of Virginia, Charlottesville, Virginia 22906, USA}
\affiliation{Waseda University, Tokyo 169, Japan}
\affiliation{Wayne State University, Detroit, Michigan 48201, USA}
\affiliation{University of Wisconsin, Madison, Wisconsin 53706, USA}
\affiliation{Yale University, New Haven, Connecticut 06520, USA}

\author{T.~Aaltonen}
\affiliation{Division of High Energy Physics, Department of Physics, University of Helsinki and Helsinki Institute of Physics, FIN-00014, Helsinki, Finland}
\author{B.~\'{A}lvarez~Gonz\'{a}lez$^z$}
\affiliation{Instituto de Fisica de Cantabria, CSIC-University of Cantabria, 39005 Santander, Spain}
\author{S.~Amerio}
\affiliation{Istituto Nazionale di Fisica Nucleare, Sezione di Padova-Trento, $^{ff}$University of Padova, I-35131 Padova, Italy}
\author{D.~Amidei}
\affiliation{University of Michigan, Ann Arbor, Michigan 48109, USA}
\author{A.~Anastassov$^x$}
\affiliation{Fermi National Accelerator Laboratory, Batavia, Illinois 60510, USA}
\author{A.~Annovi}
\affiliation{Laboratori Nazionali di Frascati, Istituto Nazionale di Fisica Nucleare, I-00044 Frascati, Italy}
\author{J.~Antos}
\affiliation{Comenius University, 842 48 Bratislava, Slovakia; Institute of Experimental Physics, 040 01 Kosice, Slovakia}
\author{G.~Apollinari}
\affiliation{Fermi National Accelerator Laboratory, Batavia, Illinois 60510, USA}
\author{J.A.~Appel}
\affiliation{Fermi National Accelerator Laboratory, Batavia, Illinois 60510, USA}
\author{T.~Arisawa}
\affiliation{Waseda University, Tokyo 169, Japan}
% ...

\collaboration{CDF Collaboration\footnote{With visitors from
$^a$Istituto Nazionale di Fisica Nucleare, Sezione di Cagliari, 09042 Monserrato (Cagliari), Italy,
$^b$University of CA Irvine, Irvine, CA 92697, USA,
$^c$University of CA Santa Barbara, Santa Barbara, CA 93106, USA,
$^d$University of CA Santa Cruz, Santa Cruz, CA 95064, USA,
$^e$Institute of Physics, Academy of Sciences of the Czech Republic, Czech Republic,
$^f$CERN, CH-1211 Geneva, Switzerland,
$^g$Cornell University, Ithaca, NY 14853, USA,
$^h$University of Cyprus, Nicosia CY-1678, Cyprus,
$^i$Office of Science, U.S. Department of Energy, Washington, DC 20585, USA,
$^j$University College Dublin, Dublin 4, Ireland,
$^k$ETH, 8092 Zurich, Switzerland,
$^l$University of Fukui, Fukui City, Fukui Prefecture, Japan 910-0017,
$^m$Universidad Iberoamericana, Mexico D.F., Mexico,
$^n$University of Iowa, Iowa City, IA 52242, USA,
$^o$Kinki University, Higashi-Osaka City, Japan 577-8502,
$^p$Kansas State University, Manhattan, KS 66506, USA,
$^q$Korea University, Seoul, 136-713, Korea,
$^r$University of Manchester, Manchester M13 9PL, United Kingdom,
$^s$Queen Mary, University of London, London, E1 4NS, United Kingdom,
$^t$University of Melbourne, Victoria 3010, Australia,
$^u$Muons, Inc., Batavia, IL 60510, USA,
$^v$Nagasaki Institute of Applied Science, Nagasaki, Japan,
$^w$National Research Nuclear University, Moscow, Russia,
$^x$Northwestern University, Evanston, IL 60208, USA,
$^y$University of Notre Dame, Notre Dame, IN 46556, USA,
$^z$Universidad de Oviedo, E-33007 Oviedo, Spain,
$^{aa}$CNRS-IN2P3, Paris, F-75205 France,
$^{bb}$Texas Tech University, Lubbock, TX 79609, USA,
$^{cc}$Universidad Tecnica Federico Santa Maria, 110v Valparaiso, Chile,
$^{dd}$Yarmouk University, Irbid 211-63, Jordan. 
}}
\noaffiliation

\begin{abstract}
A measurement of the top quark mass ($M_{\rm top}$) in the all-hadronic decay channel 
is presented. It uses 5.8\,fb$^{-1}$ of  $p\bar p$ data collected with the CDF\,II 
detector at the Fermilab Tevatron Collider.
Events with six to eight jets 
are selected by a neural network algorithm and by the requirement that 
at least one of the jets is tagged as a $b$-quark jet. 
The measurement is performed with a likelihood fit technique, 
which simultaneously determines $M_{\rm top}$ and the jet energy scale 
(${\rm JES}$) calibration.
The fit yields a value of 
\mbox{$M_{\rm top} = 172.5~\pm 1.4\,(\textnormal{stat}) ~ \pm 1.0\,({\rm JES}) ~ \pm 1.1\,(\textnormal{syst})$~GeV/$c^2$}.

\end{abstract}


\maketitle

The mass of the top quark ($M_{\rm top}$) is a fundamental parameter 
of the standard model (SM) and its large value makes the top quark contribution dominant in loop 
corrections to many observables, like the $W$ boson mass $M_W$.  
Precise measurements of $M_W$ and $M_{\rm top}$ 
allow one to set indirect constraints on the mass of the, as yet unobserved, Higgs boson~\cite{lepewg}.
\par
In this Letter we present 
a measurement of $M_{\rm top}$ using proton-antiproton collision 
events at a center-of-mass energy of 1.96\,TeV.
Top quarks are produced at the largest rate in pairs ($t\bar{t}$),
with each top quark decaying 
immediately into a $W$ boson and a $b$ quark nearly 100\% of the time\,\cite{pdg}.
In this analysis events where both the $W$'s decay to a quark-antiquark pair 
($t\bar{t} \to W^{+} b \, W^{-}\bar{b} \to q_{1}\bar{q}_{2} b \, q_{3} \bar{q}_{4} \bar{b}$)
are considered. 
This all-hadronic final state 
has the largest branching ratio among the possible decay
channels ($46\%$),
but it is overwhelmed by the QCD multijet background processes, which surpass  
$t\bar t$ production by three orders of magnitude even after  
a dedicated trigger requirement. 
Nevertheless, it will be shown how this difficult background can be successfully controlled 
and significantly suppressed with a properly optimized event selection.
%
The fundamental analysis technique is the same 
exploited to obtain the previous result 
from CDF, and is described in details in\,\cite{ahprd}. 
However, improvements in the event selection 
and a larger dataset allow us to decrease 
the total uncertainty on $M_{\rm top}$ by $21\%$.
The additional dataset has been acquired at higher instantaneous luminosity, 
which results in a higher number of background events in the data sample. 
Despite this fact, the introduction of significant improvements to the analysis results 
in the world best measurement of $M_{\rm top}$ in the all-hadronic channel so far,
also entering with the third largest weight in the $M_{\rm top}$
world average calculation\,\cite{topreview, latestCDFD0mass}.
\par
The data correspond to an integrated luminosity of $5.8$\,fb$^{-1}$. They have been collected  between 
March 2002 and February 2010 by the CDF detector,
a general-purpose apparatus designed 
to study $p\bar p$ collisions at the Tevatron and described in detail in\,\cite{CDFdetector}.   
Events used in this measurement are selected by a multijet trigger\,\cite{ahprd},  
and retained only if they are well contained 
in the detector acceptance, 
have no well identified energetic  electron or muon, 
and  have a missing transverse energy $\met$\,\cite{coordinate} 
satisfying  $\met/\sqrt{\sum E_{T}}<3$\,GeV$^\frac{1}{2}$,
where $\sum E_{T}$ is the sum of the transverse energy $E_T$ of all jets. 
Candidate events are also required to have from six to eight 
``tight'' ($E_{T} \geq 15$\,GeV and $\vert\eta\vert\le 2.0$) jets.
After this preselection, a total of about $5.7$\,M 
events is observed in the data, 
with less than 9 thousand expected from $t\bar{t}$ events.  
To improve the signal-to-background ratio ($S/B$) a $b$-tagging algorithm\,\cite{vertex} is used
to identify (``$b$-tag'' or simply ``tag'') jets 
that most likely resulted from the fragmentation of a $b$ quark.
Only events with one to three tagged jets are then retained, improving the $S/B$ by a factor of 6.
In order to further increase the signal purity, a multivariate algorithm
is implemented.  
An artificial neural network, based on a set of kinematic and jet shape variables~\cite{ahprd}, 
is used to take advantage of the distinctive features of signal and background events. 
The neural network was trained using simulated $t\bar{t}$ events generated by {\sc Pythia}\,\cite{pyt} and 
propagated through the CDF detector simulation. 
At this level of selection the fraction of signal events is still negligible so that
the data can be used to represent the background. 
The value of the output node, $N_{\rm out}$, is used as a discriminant between signal and background,
providing a gain in $S/B$ by an additional factor of about 30. 
\par
The background for the $t\bar{t}$ multijet final state comes  mainly from QCD production of 
heavy-quark pairs ($b\bar b$ and $c\bar c$) and events with false tags from light-quark and gluon jets. 
Given the large theoretical uncertainties on the QCD multijet production cross section,  
the background prediction is obtained from the data themselves.
The probability of tagging a jet in a background event ($P^{+}$)
is evaluated using data with five tight jets and passing the preselection
($S/B \approx 1/2000$). This ``tag rate'' is parametrized in terms of
a few relevant jet variables and is then used to estimate
the probability that a candidate event belongs to the background 
and contains a given number of tagged jets. As described in detail in\,\cite{ahprd} this allows to predict
the expected amount of background events in the selected samples as well as their
distributions. For example, the average number of background 1-tag events is estimated by
\begin{eqnarray}
    \sum_{\rm events} \left[ \sum_{i=1}^{N_{\rm jets}} C_{\rm 1\,tag}^{i}\cdot P^{+}_{i} \prod_{k \neq i} \left( 1 - P^{+}_{k} \right )  \right] \nonumber
\end{eqnarray}
 
where the outer sum runs over all events selected just before the $b$-tagging requirement, and the
inner one over the jets of the event. The factor $C_{\rm 1\, tag}$ represents a correction to take into account
correlations among jets within the same event\,\cite{ahprd}, and it is parametrized as a function of the same 
variables used for the tag rate. 

\par
The analysis employs the template method to measure $M_{\rm top}$ 
with simultaneous calibration of the jet energy scale ($\mathrm{JES}$)\,\cite{ahprd,2dfit},
allowing a strong reduction of the associated systematic uncertainty. 
Distributions of variables sensitive to the ``true'' values of
$M_{\rm top}$ and  ${\rm JES}$, obtained by  Monte Carlo (MC) events,
are used as a reference (``template'') in the measurement.
A maximum likelihood fit is performed to define the values
that best reproduce the same distributions as observed in the data.
An usual choice is to consider the distributions of the event-by-event 
reconstructed top quark mass, $m_{t}^{\rm rec}$,
and $W$ boson mass, $m_{W}^{\rm rec}$ as the reference templates. 
The ${\rm JES}$ is a multiplicative factor representing a correction
applied to the raw energy of a reconstructed jet ($E_{T}^{\rm raw}$),
so that its corrected energy $\Et  =  \mathrm{JES} \cdot E_{T}^{\rm raw}$,
is a better estimate of the energy of the underlying parton\,\cite{jes}.
%
Discrepancies between data and simulation result in an uncertainty
on the $\mathrm{JES}$ value to be applied in MC events to reproduce the
data, and, as a consequence, on the measurements of $M_{\rm top}$.
Nevertheless, this value can be calibrated ``in situ'', using $m_{W}^{\rm rec}$
as a template. This represents a well tested technique, first applied in\,\cite{2dfit} and
now used to obtain the most precise top quark mass measurements at the Tevatron\,\cite{topreview, latestCDFD0mass}.

The templates are built as follows\,\cite{ahprd}.
For each selected event, each of the six highest-$E_{T}$ jets is
assigned in turn to one of the six quarks  of a $t\bar{t}$ all-hadronic final state.
Then, for each combination
the jets are arranged in
two triplets (the top quarks), each including a doublet (corresponding to the $W$ boson) and a $b$ quark. 
To reduce the possible number of permutations, $b$-tagged jets are assigned to 
$b$ quarks only, resulting in 30, 6 or 18 permutations for events with one, two or three 
tagged jets, respectively\,\cite{permutations}.

For each permutation $m_t^{\rm rec}$ is obtained 
through a constrained fit based on the minimization of the following $\chi^2$-like function:
\begin{eqnarray}
\chi^{2}_{t} &  =  &  ~\frac{ \big(m_{jj}^{(1)}-M_W \big)^2 }{ \Gamma^2_W } +
                       \frac{ \big(m_{jj}^{(2)}-M_W \big)^2 }{ \Gamma^2_W }  \nonumber \\
             &     & + \frac{ \big(m_{jjb}^{(1)}-m_t^{\rm rec} \big)^2 }{ \Gamma^2_t } +
                       \frac{ \big(m_{jjb}^{(2)}-m_t^{\rm rec} \big)^2}{\Gamma^2_t}   \nonumber \\
             &     & + \sum^6_{i=1} \frac{ \big(p^{\rm fit}_{T,i}-p^{\rm meas}_{T,i} \big)^2}{\sigma^2_i} \nonumber
\end{eqnarray} 
where $m_{jj}^{(1,2)}$ are the invariant masses of the two pairs of jets assigned to light flavor quarks,
$m_{jjb}^{(1,2)}$
are the invariant masses of the triplets including one pair and one jet assigned to a $b$ quark, 
$M_{W} = 80.4$\,GeV/$c^2$ and $\Gamma_{W}=2.1$\,GeV are the measured mass and natural width 
of the $W$ boson\,\cite{pdg}, and
$\Gamma_t = 1.5$\,GeV is the assumed natural width of the top quark\,\cite{widt}. 
The jet transverse momenta are constrained in the fit to the measured values, 
$p^{\rm meas}_{T,i}$, within their known resolutions, $\sigma_i$.
The fit is performed with respect to 
$m_t^{\rm rec}$ and the transverse momenta of the jets $p^{\rm fit}_{T,i}$,
and, among all the permutations, the one which gives the lowest
value for the minimized $\chi^{2}_{t}$ is selected.
The variable $m_{W}^{\rm rec}$ is reconstructed by the same procedure considered for $m_{t}^{\rm rec}$, but with a 
$\chi^{2}$ function, $\chi^{2}_{W}$, where also the $W$ mass is left free to vary in the fit.  
The selected values of  $m_{t}^{\rm rec}$ and $m_{W}^{\rm rec}$ enter
the respective distributions, built separately for 
events with exactly  one or  $\ge 2$ tags.
\par
Signal templates are built using MC events with $M_{\rm top}$ values
from 160 to 185\,GeV/$c^{2}$, with steps of 1.0\,GeV/$c^{2}$, and, for each
value, moving the ${\rm JES}$ by $\Delta\mathrm{JES} \cdot \sigma_{\rm JES}$
from the default. Here  $\sigma_{\rm JES}$ is the absolute uncertainty on the ${\rm JES}$\,\cite{jes} and
$\Delta\mathrm{JES}$ is a dimensionless number. Values of $\Delta\mathrm{JES}$ between $-2$ and $+2$, 
in steps of $0.5$, have been used,
and in the following we refer to this parameter to denote variations
of the ${\rm JES}$. 
To construct the background templates we apply the fitting technique to the 
data events passing the neural network selection cut, omitting the $b$-tagging requirement 
(``pretag'' sample)\,\cite{ahprd}. The weight of each value 
of $m_t^{\rm rec}$ and $m_W^{\rm rec}$ is given
by the probability of the event to belong to the background and to contain tagged jets,
evaluated by the tag rates of jets, as outlined above.
\par
Sets of simulated experiments (``pseudo-experiments'', PEs) have been performed to optimize the requirements on the values 
of $N_{\rm out}$, $\chi^{2}_{t}$ and $\chi^{2}_{W}$ in order to minimize the statistical uncertainty 
on the $M_{\rm top}$ measurement. 
As an improvement with respect to\,\cite{ahprd}, two different
sets of events, denoted by  $S_{\rm JES}$ and $S_{M_{{\rm top}}}$, are used
to build the $m_{W}^{\rm rec}$ and $m_{t}^{\rm rec}$ templates, respectively.
The set $S_{\rm JES}$ is selected by using cuts on
$N_{\rm out}$ and $\chi^{2}_{W}$, while $S_{M_{{\rm top}}}$ is selected by a {\em further} requirement
on $\chi^{2}_{t}$, so that $S_{M_{{\rm top}}}$ corresponds to a subset of $S_{\rm JES}$.   
This new procedure contributes in reducing the final total uncertainty 
on $M_{\rm top}$ with respect to\,\cite{ahprd} by about $12\%$. 
%
%
Tables\,\ref{tab:1tagSel} and\,\ref{tab:2MtagSel} report the flow of the event selection for
1-tag and $\ge 2$-tag events, respectively. As the final requirements 
are optimized  separately for the two tagging categories,  
the $b$-tag requirement is included in the 
flow just after the preselection. 

\begin{table}[h!]
 \caption{
           Selection flow for 1-tag events samples.
           For each requirement the number of events observed in the data, 
           the expected number of $t\bar{t}$ signal events,
           the absolute efficiency on the
           signal ($\varepsilon$) and the signal-to-background ratio ($S/B$) are shown.
           For the signal $M_{\rm top} = 172.5$\,GeV/$c^{2}$ and $\Delta {\rm JES} = 0$ are
           used. The expectations are normalized to the integrated luminosity of the 
           data sample ($5.8$\,fb$^{-1}$) using the theoretical cross section ($7.46$\,pb),
           while the background is evaluated as the difference between the data and the expected signal.
         }
 \vspace*{-2mm}
  \begin{center}
    \begin{tabular}{lcccc}
       \hline
       \hline

 \parbox[c][6mm][c]{35mm}{Selection Requirement}    & ~Data~   &  ~$t\bar{t}$~   & ~$\varepsilon$\,($\%$)~ &   ~$S/B$~    \\
       \hline

Trigger $+$ Presel.           & \parbox[c][5.5mm][c]{13mm}{$5683210$}   &     $ 8854 $            &    $20.6$                  & $ 1/641 $  \\
 
$\equiv 1$ $b$-tag                 &  \parbox[c][5mm][c]{13mm}{$546579$}     &     $ 3861 $            &    $9.0$                   & $ 1/141 $  \\

$ N_{\rm out} > 0.97  $                 &  \parbox[c][5mm][c]{13mm}{$5743$}       &     $ 1028 $            &    $2.4$                & $ 1/4.6 $  \\

$\chi^{2}_{W} < 2 $ ($S_{\rm JES}$)  &  \parbox[c][5mm][c]{13mm}{$4368$}        &     $  881 $            &    $2.1$                   & $ 1/4.0 $  \\

$\chi^{2}_{t} < 3 $ ($S_{M_{\rm top}}$) &  \parbox[c][5mm][c]{13mm}{$2256$}        &     $  604 $            &    $1.4$                   & $ 1/2.7 $  \\                   

      \hline
      \hline
    \end{tabular}
  \end{center}

 \label{tab:1tagSel}
\end{table}

\begin{table}[h!]
  \caption{
           Selection flow for $\geq 2$-tag events samples.
           The same notations of Table\,\ref{tab:1tagSel} are used.          
         }
\vspace*{-2mm}
  \begin{center}
    \begin{tabular}{lcccc}
       \hline
       \hline

 \parbox[c][6mm][c]{35mm}{Selection Requirement} & ~Data~   & ~$t\bar{t}$~   & ~$\varepsilon$\,($\%$)~  &  ~$S/B$~   \\
   \hline

Trigger $+$ Presel. &  \parbox[c][5.5mm][c]{13mm}{$5683210$}      &     $ 8854 $            &    $20.6$                   & $ 1/641 $  \\
 
   $\geq 2$ $b$-tags     &  \parbox[c][5mm][c]{13mm}{$47229$}         &     $ 1520 $            &    $3.5$                     & $ 1/30 $  \\

   $ N_{\rm out} > 0.94  $    &  \parbox[c][5mm][c]{13mm}{$2379$}       &     $  740 $            &    $1.7$                    & $ 1/2.2 $  \\

   $\chi^{2}_{W} < 3 $ ($S_{\rm JES}$)   &  \parbox[c][5mm][c]{13mm}{$1196$}       &     $  468 $            &    $1.1$          & $ 1/1.6 $  \\
 
   $\chi^{2}_{t} < 4 $  ($S_{M_{\rm top}}$)   &  \parbox[c][5mm][c]{13mm}{$600$}       &     $  316 $            &    $0.7$         & $ 1/0.9 $  \\                   

      \hline
      \hline
    \end{tabular}
  \end{center}

 \label{tab:2MtagSel}
\end{table}

\begin{figure}[htbp!]
 \begin{center}
  \begin{tabular}{c}
    
      \epsfig{file=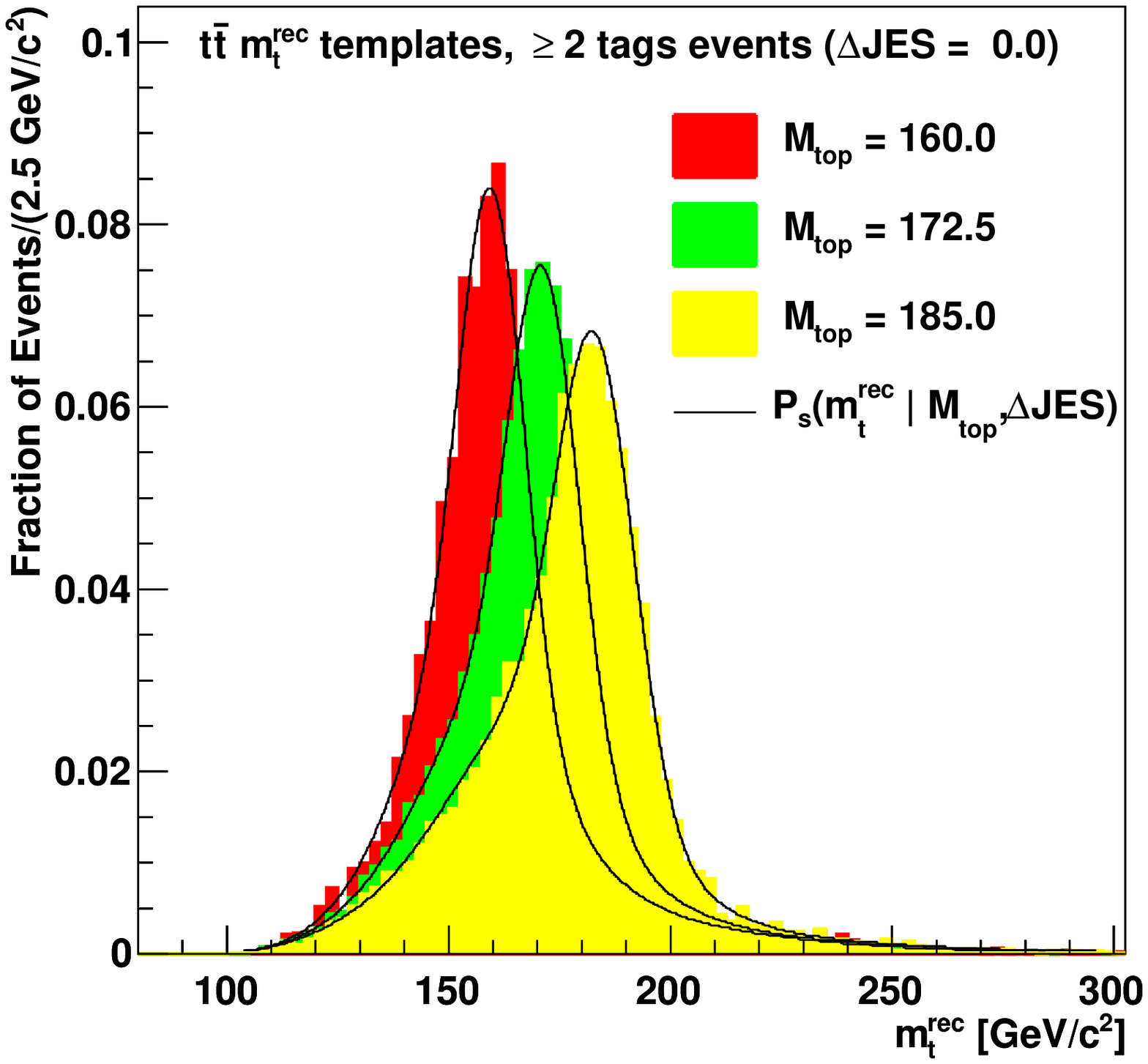,clip=true,bbllx=5pt,bblly=5pt,bburx=480pt,bbury=445pt,width=6cm}  
   \\
      \epsfig{file=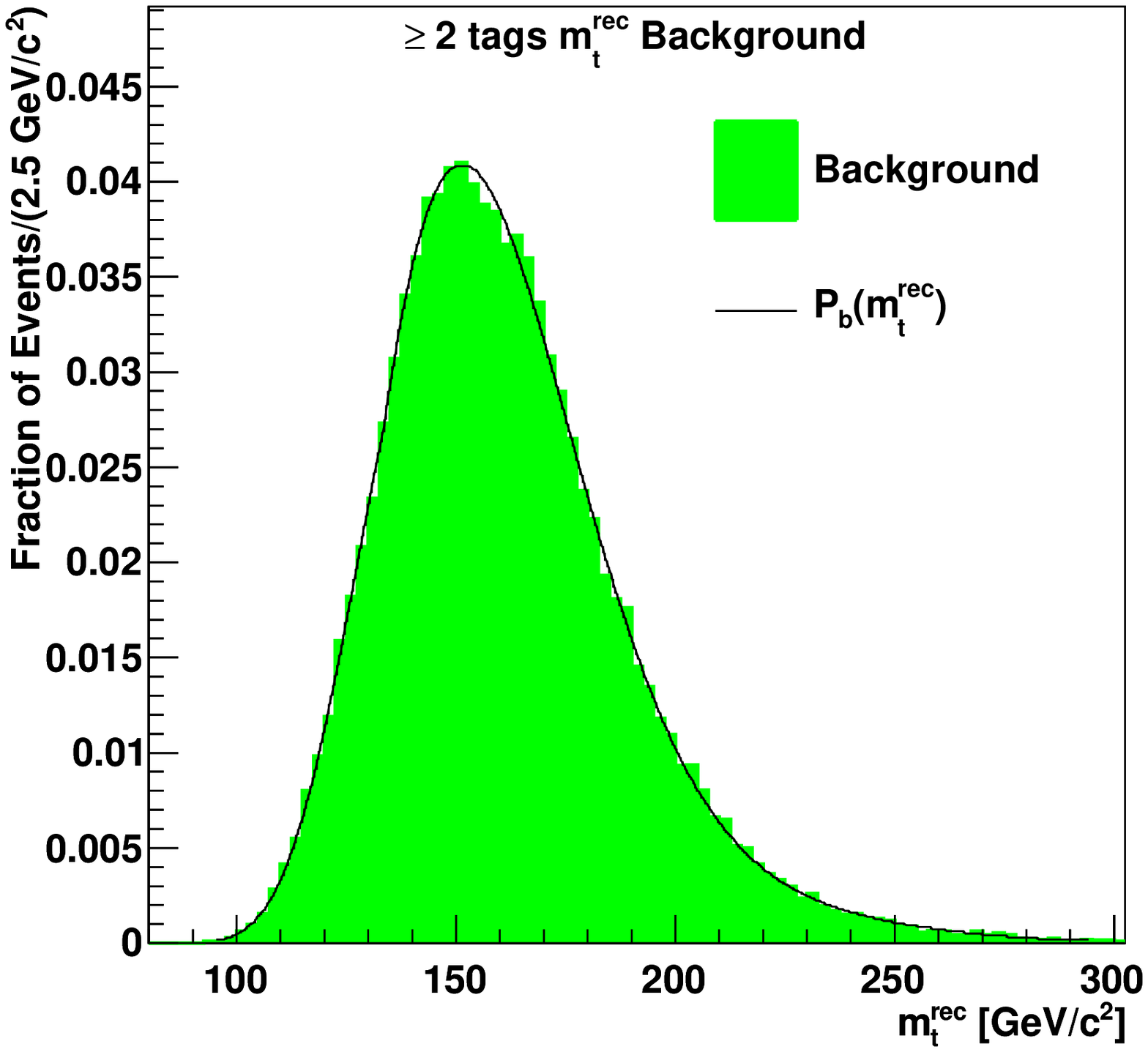,clip=true,bbllx=5pt,bblly=5pt,bburx=480pt,bbury=445pt,width=6cm}    
                          
  \end{tabular}
  \end{center}
   \vspace*{-3mm}
      \caption{
               Templates of $m_{t}^{\rm rec}$ for events with $\ge 2$ tags and corresponding
               probability density functions superimposed. Top plot: the signal p.d.f, $P_{\rm s}$, 
               for various values of $M_{\rm top}$ and $\Delta\mathrm{JES}=0$.   
               Bottom plot: the background p.d.f., $P_{\rm b}$.
              }
      \label{fig:templates}
\end{figure}

In order to measure $M_{\rm top}$ with the simultaneous calibration of the $\mathrm{JES}$, 
a fit is performed  in which an 
unbinned extended likelihood function is maximized to find the  values 
of $M_{\mathrm{top}}$, $\Delta\mathrm{JES}$, the number of 
signal ($n_{s}$) and background ($n_{b}$) 
events for each tagging category
which best reproduce the observed distributions of $m_t^{\mathrm{rec}}$ and $m_W^{\mathrm{rec}}$\,\cite{ahprd}.
The likelihood depends on the probability density functions (p.d.f.'s)
of $m_{t}^{\mathrm{rec}}$ and $m_{W}^{\mathrm{rec}}$
expected for signal ($s$) and background ($b$),
$P_{s}\left(m_{t}^{\mathrm{rec}}| \, M_{{\rm top}},\, \Delta\mathrm{JES}\right)$,
$P_{s}\left(m_{W}^{\mathrm{rec}}| \, M_{{\rm top}},\, \Delta\mathrm{JES}\right)$, 
$P_{b}\left(m_{t}^{\mathrm{rec}}\right)$, and
$P_{b}\left(m_{W}^{\mathrm{rec}}\right)$.
The notation points out that the shapes of the signal p.d.f.'s are functions 
of the fit parameters  $M_{\mathrm{top}}$ and $\Delta\mathrm{JES}$. 
This dependence is obtained by fitting the whole set of templates, initially built as histograms.
Figure\,\ref{fig:templates} shows examples of signal and background templates for the $\ge 2$-tag sample,
with the corresponding p.d.f.'s superimposed.

The presence of the different sets  $S_{\rm JES}$ and $S_{M_{{\rm top}}}$
requires the generalizations of some of the terms of
the likelihood with respect to\,\cite{ahprd}.
The function can be divided into three parts:
\begin{center}
$ {\cal L} = {\cal L}_{1\,\mathrm{tag}} \times {\cal L}_{\ge 2\,\mathrm{tags}} \times {\cal L}_{\Delta\mathrm{JES}_{\mathrm{constr}}} $
\end{center}  
where ${\cal L}_{\Delta\mathrm{JES}_{\rm constr}}$ is a Gaussian term constraining the 
$\mathrm{JES}$ to the nominal value (i.e. $\Delta\mathrm{JES}$ to $0$)
within its uncertainty\,:

\begin{eqnarray}
{\cal L}_{\Delta\mathrm{JES}_{\rm constr}} & = & 
e^{-\frac{(\mathrm{JES} - \mathrm{JES}_{\rm constr})^2}{2\sigma^2_{\mathrm{JES}}}} \nonumber \\
                                       & = & e^{-\frac{ \left[\left(\mathrm{JES}_{\rm constr} + \Delta\mathrm{JES} \cdot 
                                                           \sigma_{\mathrm{JES}}\right)   - \mathrm{JES}_{\rm constr}\right]^2}{2\sigma
^2_{\mathrm{JES}}}} \nonumber \\ 
                                       & = & e^{-\frac{\left[ \Delta\mathrm{JES} \right]^{2} }{2}} \nonumber 
\end{eqnarray}

Terms ${\cal L}_{1\,{\rm tag}}$ and ${\cal L}_{\ge 2 \, \mathrm{tags}}$ are in turn defined as:
\begin{eqnarray}
 {\cal L}_{1,\ge 2\, \mathrm{tags}} =       {\cal L}_{\Delta\mathrm{JES}} 
                                     \times {\cal L}_{M_{\mathrm{top}}} 
                                     \times {\cal L}_{\mathrm{evts}}
                                     \times {\cal L}_{N^{\mathrm{bkg}}_{\mathrm{constr}}}\,, \nonumber
\end{eqnarray}
where, omitting the dependences on $M_{\rm top}$ and $\Delta_{\rm JES}$, 
\vspace*{-1mm}
\begin{flushleft}
  $\displaystyle
    {\cal L}_{\Delta\mathrm{JES}}  = \hspace*{-1mm} \prod_{i=1}^{N_{\rm obs}^{S_{\rm JES}}} \hspace*{-1.5mm}
                             \frac{  n_s  P^{m_{W}^{\rm rec}}_{s}(m_{W,\,i})
                                   + n_b  P^{m_{W}^{\rm rec}}_{b}(m_{W,\,i})
                                  }
                                  {n_s+n_b}  
$\,,
\end{flushleft}
\vspace*{-4mm} 
 \begin{flushleft}
 $\displaystyle
              {\cal L}_{M_{\mathrm{top}}}  = \hspace*{-2mm}  \prod_{i=1}^{N_{\rm obs}^{S_{M_{{\rm top}}}}} \hspace*{-2.5mm}      \frac{ 
                         ~~{\cal A}_{s} n_s  P^{m_{t}^{\rm rec}}_{s}(m_{t,\,i}) + {\cal A}_{b} n_b P^{m_{t}^{\rm rec}}_{b}(m_{t,\,i}) 
                             }
                             {
                               {\cal A}_{s}  n_{s} + {\cal A}_{b}   n_{b}
                             } 
 $\,,
 \end{flushleft}

 \begin{eqnarray}
 \hspace*{-5mm} 
           {\cal L}_{\rm evts} & = &  \hspace*{-7mm} \sum_{   {~~~~r_{s} + r_{b} = N_{\rm obs}^{ S_{\rm JES}} \atop   }   }
                                            \hspace*{-5mm}     P(r_{s},\,n_{s}) \cdot P(r_{b},\,n_{b})
                         \cdot \nonumber \\
                             &  &  \hspace*{9mm} \left[
                                     \sum_{ 
                                                  {  t_{s} \leq r_{s},~ t_{b} \leq r_{b}
                                                   \atop
                                                     t_{s} + t_{b} = N_{\rm obs}^{S_{M_{\rm top}}}
                                                  }
                                         }
                                   \hspace*{-7mm}  B \left( t_{s},\,r_{s},\,{\cal A}_{s} \right)
                                              \cdot
                                            B \left( t_{b},\,r_{b},\,{\cal A}_{b} \right)
                             \right] \nonumber
 \end{eqnarray}

and

\begin{flushleft}
     $\displaystyle
        {\cal L}_{N^{\rm bkg}_{\rm constr}} =  e^{-\frac{\left[ n_{b}-n_{(b,\,{\rm exp})} \right] ^2}{2\sigma_{n_{(b,\,{\rm exp})}}^2}} 
     $
\end{flushleft}

In the first term the probability to observe the set $m_{W,\,i},~(i=1,...,N_{\rm obs}^{S_{\rm JES}})$ of $m_{W}^{\rm rec}$ values 
reconstructed in the data is calculated by the signal and background expected distributions, 
$P^{m_{W}^{\rm rec}}_{s}$ and $P^{m_{W}^{\rm rec}}_{b}$ respectively, 
as a function of the free parameters of the fit $M_{{\rm top}}$, $\Delta\mathrm{JES}$, $n_{s}$, and $n_{b}$.
In the second the same is done for the distributions of the observed reconstructed top masses, 
$m_{t,\,i},~(i=1,...,N_{\rm obs}^{S_{M_{\rm top}}})$,
and the $m_{t}^{\rm rec}$ probability density functions.
The factors ${\cal A}_{s}\left( M_{\rm top},\,\Delta {\rm JES}\right)$ and ${\cal A}_{b}$ 
represent the acceptance of $S_{M_{{\rm top}}}$ with  respect to  $S_{\rm JES}$
for signal and background, respectively  
(i.e., the fraction of events selected by the requirements on $\chi^{2}_{t}$ only). 
For the signal this acceptance is parametrized as a function of the fit parameters
$M_{{\rm top}}$ and $\Delta {\rm JES}$.
The third term, ${\cal L}_{\rm evts}$, gives the probability to observe simultaneously the
number of events selected in the data in the $S_{\rm JES}$ and the $S_{M_{\rm top}}$ samples,
given the assumed values for the average number of
signal ($n_{s}$) and background ($n_{b}$) events to be expected in $S_{\rm JES}$ and the acceptances
${\cal A}_{s}(M_{\rm top},\,\Delta {\rm JES})$ and ${\cal A}_{b}$. It depends on the Poisson ($P$) and Binomial ($B$) probabilities
\begin{eqnarray}
     P(r,\,n) & = & \frac{ e^{-n} \cdot n^{r} }{ r\,! } \nonumber\\
    & &                     \nonumber \\
    B\left( t,\,r,\,{\cal A} \right) & = & { r \choose t } \cdot {\cal A}^{t} \cdot (1- {\cal A})^{r-t} \nonumber
\end{eqnarray}
In the last term, ${\cal L}_{N^{\rm bkg}_{\rm constr}}$,
the parameter $n_{b}$ is constrained by a Gaussian to the {\em a priori} background estimate 
i.e. $n_{(b,\,{\rm exp})} = 3652\, \pm \, 181$ for 1-tag events and $n_{(b,\,{\rm exp})} = 718\, \pm \, 14$ for $\ge 2$-tag events.    

The possible presence of biases in  the values 
returned by the likelihood fit has been investigated.
Pseudo-experiments  are performed assuming specific
values for $M_{\rm top}$ and $\Delta\mathrm{JES}$
and ``pseudo-data'' are therefore extracted 
from the corresponding signal and background templates.
The results of these PEs have been compared to the input values, and 
calibration functions to be applied to the output from the fit
have been defined in order to obtain, on 
average, a more reliable estimate of the true values and uncertainties. 
\par
Finally, the likelihood fit is applied to data.
After the event selection described above, we are left 
with 4368 and 1196 events with one and $\ge 2$ tags (147 have 3 tags), 
respectively, in the $S_{\rm JES}$ sample.
The corresponding expected backgrounds amount to 
$3652 \pm 181$ and $718 \pm 14$ events, respectively.
The tighter requirements used for the $S_{M_{{\rm top}}}$ samples select 2256 with one tag and 600 with $\ge 2$ tags
(76 have 3 tags),
with average background estimates of  $1712 \pm 77$ and $305 \pm 22$ events, respectively.
\par
\begin{figure}[htbp!]
\begin{center}
  \begin{tabular}{c}  

      \epsfig{file=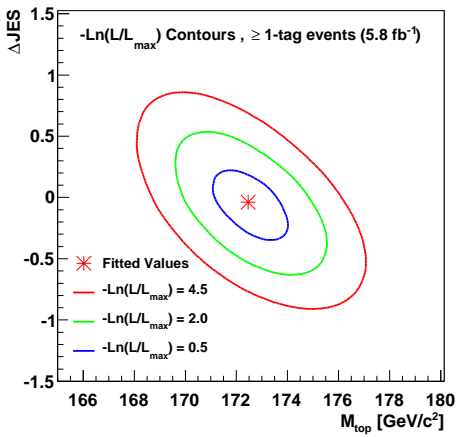,clip=true,bbllx=10pt,bblly=25pt,bburx=382pt,bbury=375pt,width=6cm}

  \end{tabular} 
  \end{center}
  \vspace*{-3mm}
  \caption{ Negative log-likelihood contours for the likelihood fit performed for the 
            $M_{\rm top}$ and  $\Delta \mathrm{JES}$ measurement. 
             The minimum is shown along with the contours whose projections correspond to one, two, and three $\sigma$ 
             uncertainties on the $M_{\rm top}$ and $\Delta \mathrm{JES}$ measurements.
           }
\label{fig:contour}
\end{figure}

For these events the variables $m_{W}^{\rm rec}$ and $m_{t}^{\rm rec}$ have been reconstructed 
and used as the data inputs to the likelihood fit.
Once the calibration procedure has been applied, the 
measurements of $M_{\rm top}$ and $\Delta \mathrm{JES}$ are
\begin{eqnarray}
 M_{\rm top} & = &  172.5 \pm 1.4\,(\textnormal{stat}) \pm 1.0\, (\mathrm{JES})~\mathrm{GeV}/c^{2}  \nonumber \\
\Delta \mathrm{JES} & = &  -0.1  \pm 0.3\,(\textnormal{stat})  \pm 0.3\, (M_{\rm top}) \,.\nonumber
\end{eqnarray}
Figure\,\ref{fig:contour} shows the measured values together with the negative log-likelihood 
contours whose projections correspond to one, two, and three $\sigma$ uncertainties on the values 
of $M_{\rm top}$ and $\Delta \mathrm{JES}$ as obtained from the likelihood fit.

\begin{figure}[t!]
  \begin{center}
        \begin{tabular}{c}

	 \epsfig{file=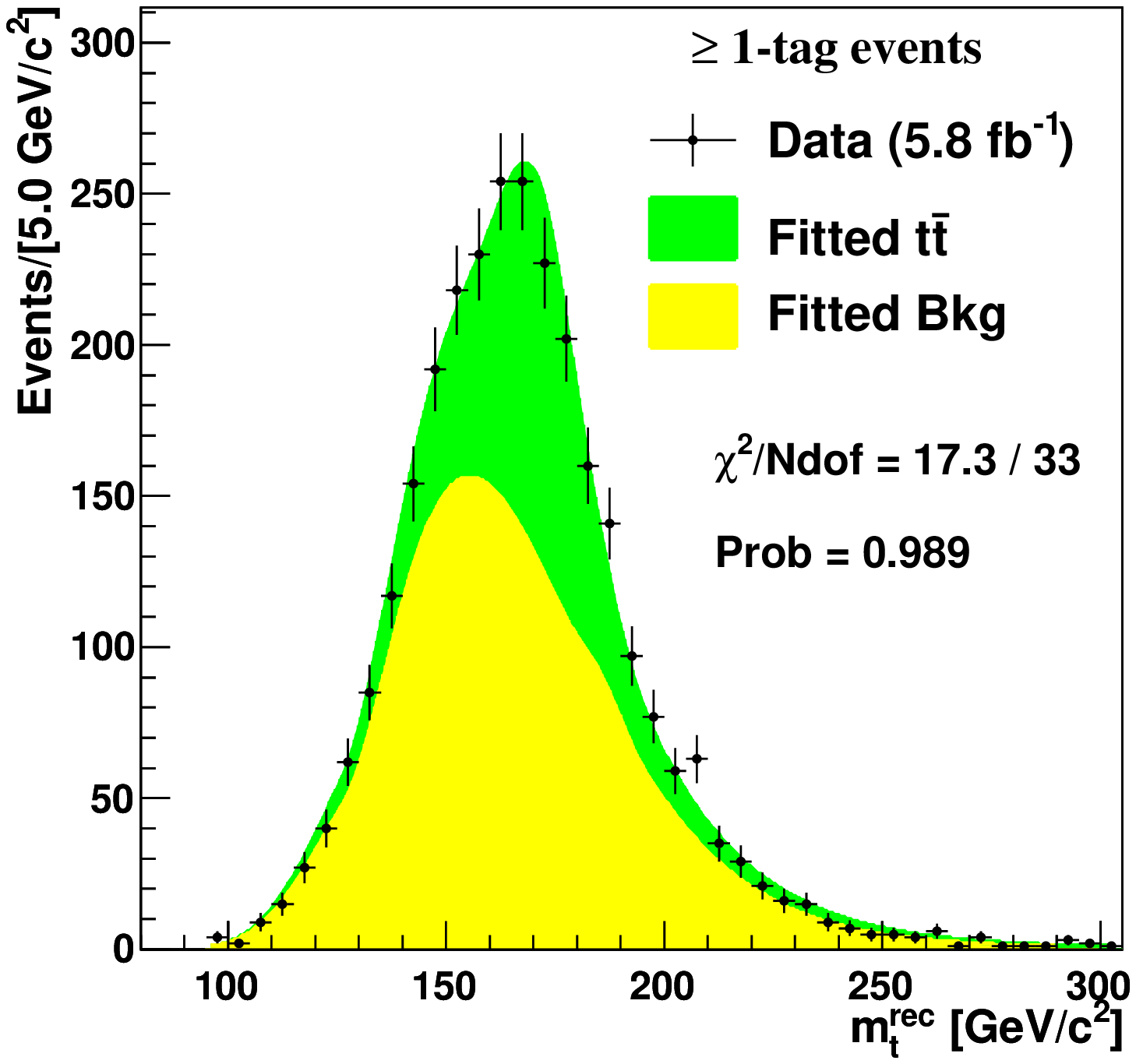,clip=true,bbllx=5pt,bblly=5pt,bburx=425pt,bbury=405pt,width=6cm} 
   
         \\   
	   
         \epsfig{file=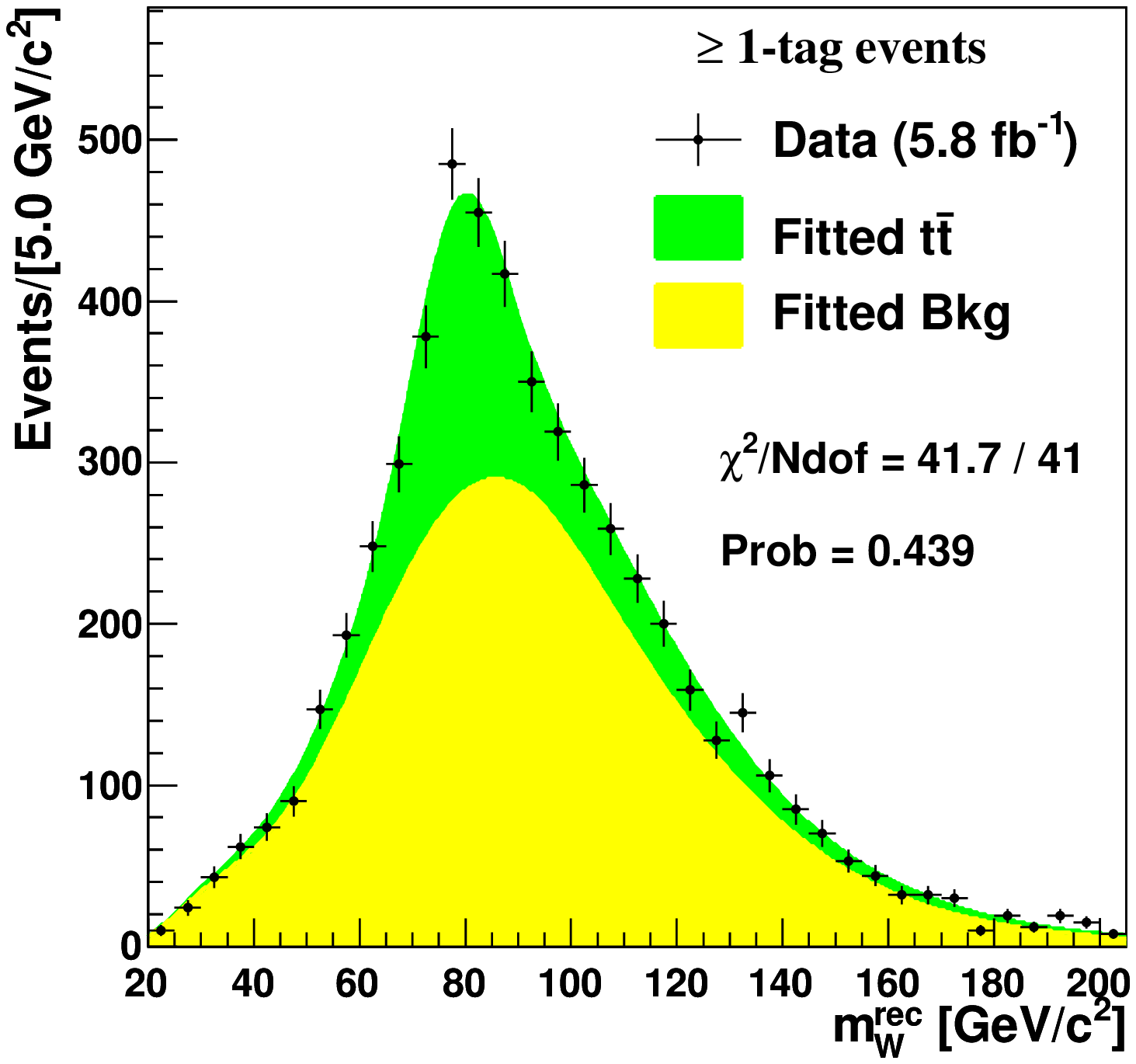,clip=true,bbllx=5pt,bblly=20pt,bburx=425pt,bbury=405pt,width=6cm}
   \end{tabular}
   \end{center} 
  \vspace*{-3mm}
   \caption{
            Distributions of $m_{t}^{\rm rec}$ (top plot) and $m_{W}^{\rm rec}$ (bottom plot) 
            as obtained in the selected data (black points) with $\ge 1$ tags,
            compared to the distributions from signal and background corresponding to the measured values of 
            $M_{\rm top}$ and $\Delta\mathrm{JES}$. 
            The expected distributions are 
            normalized to the best fit yields. 
           }
   \label{fig:topmeas}
 \end{figure}

Figure~\ref{fig:topmeas} shows the $m_{t}^{\rm rec}$ and $m_{W}^{\rm rec}$ 
distributions for the data compared to the expected background and the signal for $M_{\rm top}$ 
and $\Delta \mathrm{JES}$ corresponding to the measured values.   
The signal and background distributions are normalized to the respective yields as fitted to the data,
with the 1-tag and $\geq 2$-tag contributions summed together.

Various sources of systematic uncertainties affect the $M_{\rm top}$ and $\Delta\mathrm{JES}$  
measurements, as described in\,\cite{ahprd}. 
They are evaluated by performing PEs using 
templates built by signal samples where effects due to
systematic uncertainties have been included.
The  differences in the average values of $M_{\rm top}$ and $\Delta\mathrm{JES}$ with respect
to the PEs performed with default templates are then taken into account.
Possible residual biases existing after the calibration, and uncertainties on the parameters
of the calibration functions are also taken into account. 
The largest contributions come from uncertainties on the modeling of the background,
on the simulation of $t\bar{t}$ events, and on the individual corrections 
which $\mathrm{JES}$ depends on\,\cite{jes}.  
Table~\ref{tab:obssyst} shows a summary of all the systematic uncertainties. 
\begin{table}[tb!]
\caption{
          Sources of systematic uncertainty 
          affecting the $M_{\rm top}$ and $\Delta \mathrm{JES}$ measurements.
          The total uncertainty is obtained by the quadrature sum of each contribution. 
         }
\begin{center}
  \begin{tabular}{lccc}
  \hline
  \hline

  \parbox[c][5mm][c]{6mm}{Source} &~$\delta M_{\rm top}$~   & ~$\delta \Delta \mathrm{JES}$~ \\
                                   &     ~(GeV/$c^2$)~      &                                 \\
  \hline

  Residual bias                           &  \parbox[c][4.0mm][c]{5mm}{$ 0.2 $}   & $ 0.03 $   \\
  Calibration                             &  \parbox[c][3.5mm][c]{5mm}{$ 0.1 $}   & $ 0.01 $   \\
  Generator                               &  \parbox[c][3.5mm][c]{5mm}{$ 0.5 $}   & $ 0.21 $   \\ 
  Initial\,/\,final state radiation       &  \parbox[c][3.5mm][c]{5mm}{$ 0.1 $}   & $ 0.04 $   \\
  $b$-jet energy scale                    &  \parbox[c][3.5mm][c]{5mm}{$ 0.2 $}   & $ 0.05 $   \\ 
  $b$-tag                                 &  \parbox[c][3.5mm][c]{5mm}{$ 0.1 $}   & $ 0.01 $   \\ 
  Residual $\mathrm{JES}$                 &  \parbox[c][3.5mm][c]{5mm}{$ 0.4 $}   & $  --  $   \\
  Parton distribution functions           &  \parbox[c][3.5mm][c]{5mm}{$ 0.2 $}   & $ 0.04 $   \\
  Multiple $p\bar p$ interactions         &  \parbox[c][3.5mm][c]{5mm}{$ 0.1 $}   & $ 0.04 $   \\
  Color reconnection                      &  \parbox[c][3.5mm][c]{5mm}{$ 0.3 $}   & $ 0.12 $   \\
  Statistics of templates                 &  \parbox[c][3.5mm][c]{5mm}{$ 0.3 $}   & $ 0.05 $   \\
  Background                              &  \parbox[c][3.5mm][c]{5mm}{$ 0.6 $}   & $ 0.11 $   \\ 
  Trigger                                 &  \parbox[c][3.5mm][c]{5mm}{$ 0.2 $}   & $ 0.04 $   \\ 

  \hline

  Total                                   &  \parbox[c][5.0mm][c]{5mm}{$ 1.1 $}   & $ 0.29 $   \\ 

  \hline
  \hline
  \end{tabular}
\end{center}

\label{tab:obssyst}
\end{table} 

In summary, we have presented a measurement of the top quark mass in the all-hadronic channel,
using $p\bar{p}$ collision data corresponding to an integrated luminosity
of $5.8$\,fb$^{-1}$.
%
%
An optimized event selection, based mainly on a neural network and a $b$-tagging algorithm,
allows us to select candidate event samples with $S/B$ close to 1 in spite of the huge
background still existing at trigger level. 
The simultaneous calibration of the jet energy scale, following a well
established technique, allows to reduce down to 1\,GeV/$c^{2}$ the systematic uncertainty
due to this source.
The value obtained for the ${\rm JES}$ is in agreement both with the default value\,\cite{jes} 
and with the results obtained by other measurements of the top quark mass performed by the 
CDF Collaboration using the {\em in situ} calibration technique\,\cite{topreview, latestCDFD0mass}.   
The measured value of the top quark mass is 
$M_{\rm top}=172.5~\pm 1.4\,({\rm stat})~ \pm 1.0\,({\rm JES})~\pm 1.1\,({\rm syst})$ GeV/$c^2$,
with a total uncertainty of $2.0$\,GeV/$c^{2}$. 
This result complements and is consistent with the most recent measurements obtained in 
other channels by the CDF\,and D0 Collaborations, and also
represents the most accurate all-hadronic
measurement at the Tevatron so far. 
%

We thank the Fermilab staff and the technical staffs of the participating institutions for their vital contributions. 
This work was supported by the U.S. Department of Energy and National Science Foundation; 
the Italian Istituto Nazionale di Fisica Nucleare; the Ministry of Education, Culture, Sports, Science and Technology of Japan; 
the Natural Sciences and Engineering Research Council of Canada; the National Science Council of the Republic of China; 
the Swiss National Science Foundation; the A.P. Sloan Foundation; the Bundesministerium f\"ur Bildung und Forschung, Germany; 
the Korean World Class University Program, the National Research Foundation of Korea; 
the Science and Technology Facilities Council and the Royal Society, UK; the Russian Foundation for Basic Research; 
the Ministerio de Ciencia e Innovaci\'{o}n, and Programa Consolider-Ingenio 2010, Spain; 
the Slovak R\&D Agency; the Academy of Finland; and the Australian Research Council (ARC).


\begin{thebibliography}{99}

\bibitem{lepewg} The ALEPH, CDF, D0, DELPHI, L3, OPAL, SLD Collaborations,
                 The LEP Electroweak Working Group, the Tevatron  Electroweak Working Group, 
                 and the SLD electroweak and heavy flavour groups,
                 {\it Precision Electroweak Measurements and Constraints on the Standard Model}, 
                  arXiv:1012.2367 (2010). 

\bibitem{pdg}  K.\,Nakamura {\it et al.} (Particle Data Group), J.\,Phys. G {\bf 37}, 075021 (2010).


\bibitem{ahprd} T.\,Aaltonen {\it et al.} (CDF Collaboration), 
                Phys.\,Rev.\,D {\bf 81} (2010) 052011.

\bibitem{topreview} A.\,Barbaro Galtieri, F.\,Margaroli, I.\,Volobouev,
                    {\it Precision measurements of the top quark mass from the Tevatron in the pre-LHC era},
                    to be published in Reports on Progress in Physiscs. 
                    See also arXiv:1109.2163 [hep-ex] (2011). 

\bibitem{latestCDFD0mass} The Tevatron Electroweak Working Group for the  CDF and D0 Collaborations,
                          {\it Combination of CDF and D0  results on the mass of the top quark 
                               using up to $5.8$\,fb$^{-1}$ of data},
                          arXiv:1107.5255 [hep-ex] (2011).

\bibitem{CDFdetector} D.\,Acosta {\it et al.} (CDF Collaboration), Phys. Rev. D {\bf 71} (2005) 032001.

\bibitem{coordinate} We use a cylindrical coordinate system where $\theta$ is the polar angle 
                     with respect to the
                     proton beam direction ($z$ axis), 
                     $\phi$ is the azimuthal angle about the beam axis, 
                     and the
                     pseudorapidity is defined as $\eta = - \ln \tan(\theta/2)$.
                     A particle's transverse momentum $p_{T}$ and tranverse energy
                     $E_{T}$ are given by $|p|\,\sin\theta$  and $E\, \sin\theta$  respectively.
                     The missing $E_{T}$ vector, $\vec{\met}$, is defined by 
                     $\vec{\met} = -\sum_{i}\,E_{T,i} \,\hat{n}_{T,i}$ where $\hat{n}_{T,i}$
                     is the unit vector in the $x - y$ plane pointing from the primary interaction
                     vertex to a given calorimeter tower $i$, 
                     and $E_{T,i}$ is the $E_{T}$ measured in that tower.
                     Finally $\met = |\vec{\met}|$. 


\bibitem{vertex}  D.\,Acosta {\it et al.} (CDF Collaboration), Phys.\,Rev.\,D {\bf 71} (2005) 052003.                     

\bibitem{pyt} T.\,Sj\"{o}strand {\it et al.}, Comput.\,Phys.\,Commun. {\bf 135} (2001) 238.

\bibitem{2dfit}  A.\,Abulencia {\it et al.} (CDF Collaboration), Phys.\,Rev.\,D {\bf 73} (2006) 032003. 

\bibitem{jes}  A.\,Bhatti {\it et al.}, Nucl.\,Instrum.\,Methods Phys.\,Res., Sect. A {\bf 566} (2006) 375.


\bibitem{permutations} If three $b$-tagged jets are present in the event, the three possible
                       assignments of two out of three of them to $b$ quarks are also considered, while the
                       remaining one is treated as a light flavor jet.

\bibitem{widt} S.\,M.\,Oliveira, L.\,Brucher, R.\,Santos and A.\,Barroso,  Phys.\,Rev. D {\bf 64}, 017301 (2001).


\end{thebibliography}
\end{document}